\documentclass[twocolumn,10pt]{tsfp}
\usepackage{flushend}
\usepackage{graphicx}
\usepackage[authoryear,round]{natbib}
\usepackage{fancyhdr}
\usepackage{amsmath}
\usepackage[hidelinks,colorlinks,allcolors=blue]{hyperref}
 
\title{Direct numerical simulations of shocklet-containing turbulent channel counter-flows}

\renewcommand{\expauthors}{1.1}

\author{Arash Hamzehloo
    \affiliation{Turbulence Simulation Group\\
	Department of Aeronautics\\
	Imperial College London\\
	London SW7 2AZ, UK\\
    a.hamzehloo@imperial.ac.uk
    }	
}

\author{David J. Lusher
          \affiliation{Aerodynamics and Flight Mechanics Group\\
	University of Southampton\\
	Southampton SO16 7QF, UK\\
	d.lusher@soton.ac.uk
	
    }
}

\author{Sylvain Laizet
    \affiliation{Turbulence Simulation Group\\
	Department of Aeronautics\\
	Imperial College London\\
	London SW7 2AZ, UK\\
	s.laizet@imperial.ac.uk
    }	
}

\author{Neil D. Sandham
          \affiliation{Aerodynamics and Flight Mechanics Group\\
	University of Southampton\\
	Southampton SO16 7QF, UK\\
	n.sandham@soton.ac.uk
    }
}

\begin{document}

\maketitle   
\thispagestyle{fancy}

\fontsize{9}{11}\selectfont

\section*{ABSTRACT}
Counter-flow or counter-current configurations can maintain high turbulence intensities and exhibit a significant level of mixing. We have previously introduced a wall-bounded counter-flow turbulent channel configuration (Physical Review Fluids, 6(9), p.094603.) as an efficient framework to study compressibility effects on turbulence. Here, we extend our previous direct numerical simulation study to a relatively higher Mach number ($M=0.7$) to investigate strong compressibility effects (also by reducing the Prandtl number from $Pr=0.7$ to $0.2$), and the formation and evolution of unsteady shocklet structures. It is found that the configuration is able to produce highly turbulent flows with embedded shocklets and significant asymmetry in probability density functions of dilatation. A peak turbulent Mach number close to unity is obtained, for which the contribution of the dilatational dissipation to total dissipation is nevertheless found to be limited to 6\%. 
  
\section*{INTRODUCTION}

Traditionally, free shear layers (such as jets, wakes and mixing layers) and Poiseuille/Couette type flows (such as channel flows) have been utilized to study compressible turbulence \citep{freund2000compressibility}. Spatially-developing mixing layer simulations are computationally expensive and sensitive to far-field and inflow/outflow boundary conditions, but the basic compressibility effects are captured in temporal simulations \citep{vreman1996compressible}. A drawback of such simulations is that the shear layer thickens as time progresses and large structures swiftly fill the domain. On the other hand, Poiseuille/Couette type flows are relatively efficient to compute and can achieve high Reynolds numbers for relatively low computational costs. However, they are limited in terms of the turbulent Mach number that can be achieved. 
\par
Counter-flows are highly efficient mixers due to the high turbulence intensities that can be maintained \citep{humphrey1981tilting, strykowski1993mixing, forliti2005experimental}. Previously, we introduced \citep{hamzehloo2021direct} a wall-bounded counter-flow turbulent channel configuration, amenable to Direct Numerical Simulation (DNS), and demonstrated that it could overcome the above-mentioned barriers associated with free shear layers and Poiseuille/Couette type flows. Specifically, it retains a statistically stationary one-dimensional solution, in common with conventional channel flows, but contains an inflectional mean flow, representative of free shear layers. The counter-flow channel has periodic streamwise and spanwise boundaries and isothermal no-slip walls and is driven by a mean pressure gradient introduced by a hyperbolic tangent forcing term. It was shown that when the peak local mean Mach number reached $\sim0.55$, a turbulent Mach number of $\sim 0.6$ could be obtained, indicating that such flow configuration could potentially be useful for studying compressibility effects on turbulence.     
\par
Our earlier study \citep{hamzehloo2021direct} was limited to a maximum Mach number of $M=0.4$ (defined based on a reference velocity deduced from the forcing as discussed in the next section) since higher values necessitated using a shock-capturing method. Here, by applying an appropriate numerical treatment for flow discontinuities, we extend our previous DNS study to a Mach number of $M=0.7$. This help us investigate, for the first time in such acounter-flow configuration, high compressibility effects, and the formation and evolution of shocklets, as well as  their interactions with turbulence.

\begin{figure*}[!h]
\centering
\includegraphics[width=3.6in]{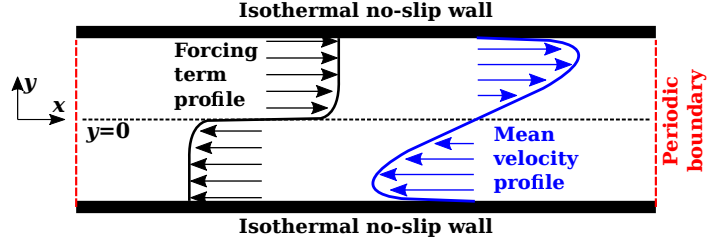}
\caption{2D schematic of the 3D counter-flow channel configuration.}
\label{figure1}
\end{figure*}

\begin{table*}
\caption{DNS counter-flow channel cases.}
\begin{tabular}{c c c c c c c c c c c c c}
\hline
 Case  &$M$ &$Re$& $Pr$&WENO Filter & $\Delta$t& $\{ u \}_{b}$&$\{ u \}_{p}$& $\langle a_c \rangle_{b}$& $\langle a_c \rangle_{p}$&$\langle T \rangle_{p}$& $\langle M \rangle_{p}$&$M_{t_{p}}$\\
\hline
  1  &0.1 &400 & 0.7& No& $2\times 10^{-4}$&1.409&2.159&10.291&10.479& 1.098&0.207& 0.213\\
  2  &0.4 &400 & 0.7& No&$5\times 10^{-5}$&1.375&2.047&3.669&3.849& 2.374&0.556& 0.595\\
  3  &0.4 &400 & 0.7& Yes& $5\times 10^{-5}$&1.380&2.062& 3.671&3.852&2.376&0.560& 0.594\\
  4  &0.7 &400 & 0.7& Yes& $5\times 10^{-5}$&1.358&2.005&2.919&3.132& 4.823&0.697& 0.758\\
  5  &0.7 &400 & 0.2& Yes& $5\times 10^{-5}$&1.535&2.238& 2.382&2.602&3.332&0.973& 0.981\\
\hline
\end{tabular}
\label{table_cases}
\end{table*}

\section*{METHODOLOGY}

\subsection*{Computational Approach}

The dimensionless governing equations of a compressible Newtonian fluid flow that conserve mass, momentum and energy are solved \citep{hamzehloo2021direct}. The latter two equations containing the shear-forcing term are given as:

\begin{align} \label{eq1}
\begin{split}
&\frac{\partial \rho u_i} {\partial t} + \frac{\partial } {\partial x_j} (\rho u_i u_j + p \delta_{ij} - \tau_{ij}) + c_j\delta_{ij}= 0, \\
&\frac{\partial \rho E} {\partial t} + \frac{\partial } {\partial x_j} (\rho E u_j + u_j p + q_j - u_i \tau_{ij}) + c_j u_j= 0,
\end{split}
\end{align}

\noindent where $\rho$ represents the density, $u_{i}(i=1,2,3)$ denotes the velocity component ($u$, $v$ and $w$, respectively) in the $i^{th}$ direction ($x$, $y$ and $z$, respectively), $E$ is the total energy, and $p$ and $\delta$ denote the pressure and the Kronecker delta, respectively. The forcing term $c_{j}$ drives the flow, with a value of $c_1=-c_0\tanh(a y)$ in the $x$ direction and zero in other directions. The maximum value is set as $c_0=1$. Since $-1H<y<1H$, where $H=1$ is the channel half height here, equivalent driving forces are applied in opposite directions to the upper and lower halves of the domain which consequently result in the formation of a shear-forcing or counter-flow condition. The coefficient $a$ in the forcing term is positive  with a value of $a=100$. The viscous stress tensor ($\tau_{ij}$) and the heat flux ($q_j$) are defined as $ \displaystyle \tau_{ij} = \frac{\mu}{Re} \bigg( \frac{\partial u_i} {\partial x_j} + \frac{\partial  u_j} {\partial x_i} - \frac{2}{3} \frac{\partial u_k}{\partial x_k} \delta_{ij}\bigg)$ and $ \displaystyle q_j = \frac{-\mu}{(\gamma -1){M}^2 Pr Re} \frac{\partial T}{\partial x_j}$, respectively. Here, $\mu$ denotes the dynamic viscosity, $T$ is the temperature, $\gamma$ is the ratio of specific heats with a value of $\gamma =1.4$ here and $Pr$ is the Prandtl number. $Re$ and $M$ denote the Reynolds and Mach numbers based on a reference velocity deduced from the forcing as $\displaystyle u_{\rm ref}^\ast=\sqrt{c_0 H/\langle\rho\rangle_b}$, where $\langle\rho\rangle_b$ is the bulk-averaged density, together with the channel half height and the wall temperature and viscosity. Angle brackets $\langle \rangle$ denote averages over the homogeneous spatial directions ($x$ and $z$) and time. The additional subscript $b$ here denotes an additional average over $y$. The dynamic viscosity is calculated as $\mu = T^{0.7}$. The temperature is calculated as $T= p\gamma {M}^2/ \rho$. Here the pressure of an ideal Newtonian fluid is obtained using an equation of state as $\displaystyle p = (\gamma -1) (\rho E - \frac{1}{2} \rho u_i u_i)$.
\par
A fourth order finite-difference central scheme is used to discretise the equations recasted in split skew-symmetric formulations to improve stability. A Lax-Friedrichs Weighted Essentially Non-Oscillatory (WENO) filter \citep{YEE2018331} is applied to the flow field after the completion of each time step to improve the solution stability in the presence of shocklets. The density field is corrected if necessary, after applying the filtering to maintain the conservation of mass. A low-storage three-stage explicit Runge-Kutta scheme is used to advance the solution in time. The simulations are performed with the flow solver OpenSBLI \citep{lusher2021opensbli}. 

\subsection*{Problem Specifications}

As shown schematically in figure \ref{figure1}, the streamwise and spanwise boundaries of the counter-flow channel configuration are periodic, while isothermal ($T_w=1.0$) no-slip walls are assigned to the boundaries in the normal direction ($y$). In order to accurately resolve the near wall region, the grid is stretched in the $y$ direction. We have previously identified an optimum domain size of $12H \times 2H \times 6H$ with a grid resolution of $240 \times 151 \times 200$ \citep{hamzehloo2021direct}. This domain size is used with a Reynolds number of $Re=400$. Mach number values of $M=0.1, 0.4$ and $0.7$ are examined. The Prandtl number has a value of mainly $Pr=0.7$. However, a case with $Pr=0.2$ ($M=0.7$) is also studied. This reduction in $Pr$ increases the the wall heat transfer, reducing the bulk temperature and sound speed in the channel, hence increasing the compressibility effect. The previously-mentioned WENO filtering is necessary to obtain a stable solution for the cases with $M=0.7$. However, for the case with $M=0.4$, results of two simulations without and with the filtering are provided to make direct comparisons and examine the filtering effect. A list of test cases studied here is presented in table \ref{table_cases}. It should be noted that, here, subscript $p$ denotes the peak values of flow quantities.

\par
Here, the single prime $^{\prime}$ denotes the turbulent fluctuation which for an arbitrary flow quantity ($\phi$) is defined as $\phi^{\prime}=\phi-\langle \phi\rangle$. Moreover, for the higher Mach number case, the Favre average is defined as $\{ \phi\}=\langle \rho \phi \rangle/\langle \rho \rangle$ and the double prime $^{\prime\prime}$ denotes the turbulent fluctuation with respect to the Favre average defined as $\phi^{\prime\prime}=\phi-\{ \phi\}$. For the Reynolds stresses, the Favre average is related to the Reynolds average as $\langle \rho \rangle \{u_i^{\prime\prime}u_j^{\prime\prime}\}=\langle \rho u_i u_j \rangle - \langle \rho \rangle \langle u_i \rangle \langle u_j \rangle$. Also, the mean Mach number is defined as $\langle M \rangle = {\sqrt{\langle u \rangle^2+\langle v \rangle^2+\langle w \rangle^2}}/{{\langle a_c \rangle}}$, where $a_c$ is the local speed of sound, while the turbulent Mach number is defined as $M_t = {\sqrt{\langle u^{\prime}u^{\prime} \rangle+\langle v^{\prime}v^{\prime} \rangle+\langle w^{\prime}w^{\prime} \rangle}}/{{\langle a_c \rangle}}$.

\section*{RESULTS AND DISCUSSION}
\subsection*{Mean Flow And Turbulence Statistics}

\begin{figure*}
\centering
\includegraphics[width=4.95in]{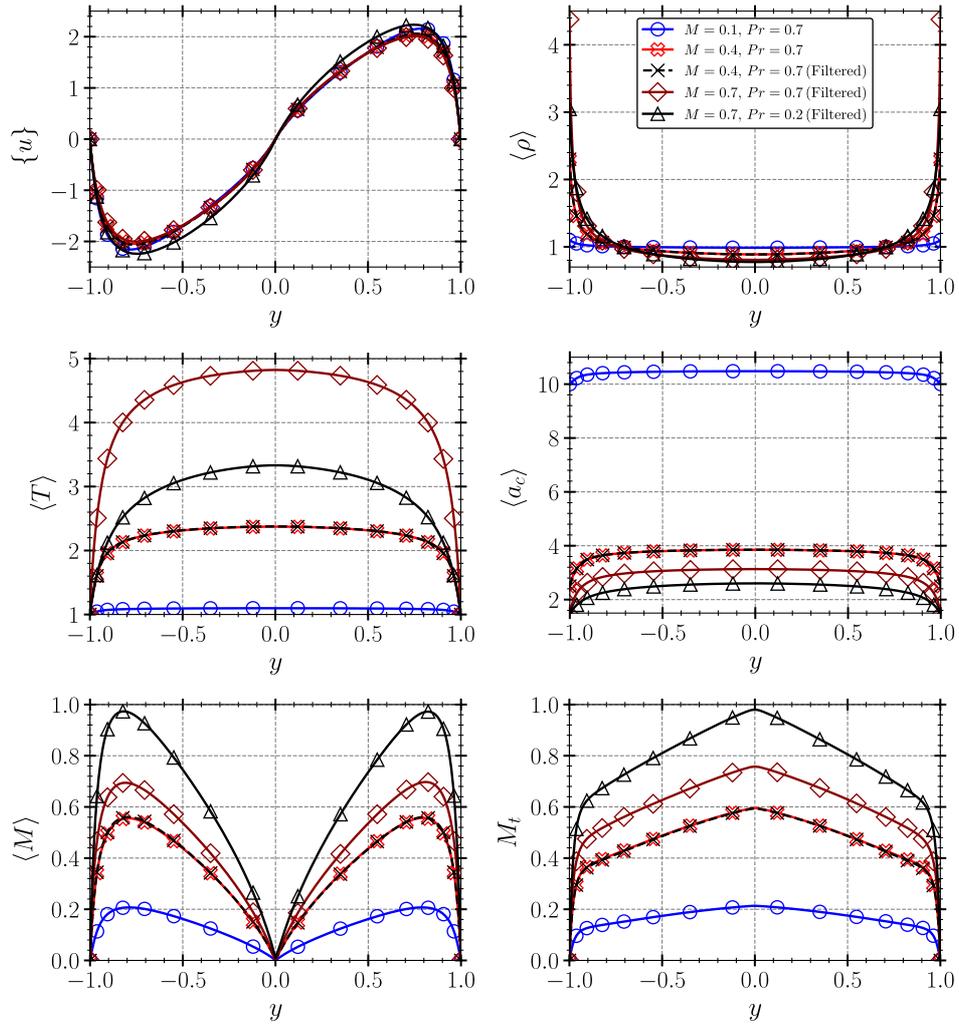}
\caption{\label{fig_mean} Profiles of the mean velocity, density, temperature, speed of sound and Mach number, and profile of the turbulent Mach number ($M_t$).}
\end{figure*}

\begin{figure*}[!ht]
\centering
\includegraphics[width=4.95in]{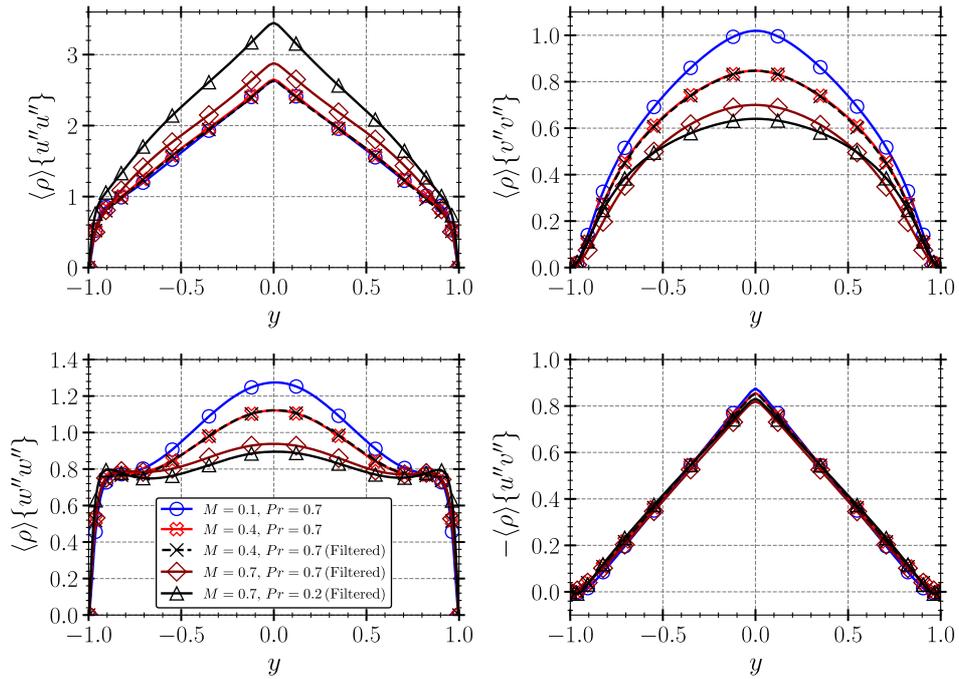}
\caption{\label{fig_stress} Profiles of the Favre Reynolds stresses.}
\end{figure*}

Figure \ref{fig_mean} shows a direct comparison between the counter-flows studied here based on various mean flow quantities, including the streamwise velocity $\{ u \}$, density $\langle \rho \rangle$, temperature $\langle T \rangle$, local speed of sound $\langle a_c \rangle$ and Mach number $\langle M \rangle$, and also the turbulent Mach number $M_t$. Additionally, figure \ref{fig_stress} provides the Favre Reynolds stresses of the counter-flows. The flow significantly heats up when the Mach number increases, for instance, as also provided in table \ref{table_cases}, the peak mean temperature is $\sim 2.0$ and $\sim 4.4$ times higher for the cases with $M=0.4$ and $M=0.7$ compared to the case with $M=0.1$, respectively. However, reducing the Prandtl number from $Pr=0.7$ to $0.2$ for $M=0.7$, reduces the peak mean temperature by $\sim 31\%$. As shown in figure \ref{fig_mean}, the mean velocity does not change significantly when changing the Mach number since the latter is governed by the local speed of sound. Mean and turbulent Mach numbers exhibit significant surges when $M$ increases. Specifcially, the peak turbulent Mach number increases by $\sim 179\%$ and $\sim 256\%$ when the Mach number increases from $M=0.1$ to $M=0.4$ and $0.7$, respectively. Prandtl number reduction while keeping the Mach number constant at with $M=0.7$, further increases the turbulent Mach number by $\sim 29.5 \%$ helping it reach a value of near unity at the channel centreline.     
\par
With respect to the normal stresses, as shown in figure \ref{fig_stress}, by increasing the compressibility through increasing the Mach number and/or reducing the Prandtl number, the peak streamwise stress increases and the peaks of the other two normal stresses reduce. For instance, the peak $\langle \rho \rangle \{ u^{\prime\prime}u^{\prime\prime} \}$ increases by around $\sim 9.5 \%$ when the Mach number increases from $M=0.4$ to $0.7$. Reducing the Prandtl number while keeping the Mach number constant at $M=0.7$ shows a noticeable effect on the normal stresses particularly the normal streamwise stress. Specifically, reducing the Prandtl number from $Pr=0.7$ to $0.2$ increases the the peak $\langle \rho \rangle \{ u^{\prime\prime}u^{\prime\prime} \}$ stress by $\sim 20\%$. However, the Favre shear stress $\langle \rho \rangle \{ u^{\prime\prime}v^{\prime\prime} \}$ continue varying linearly with $y$ as the compressibility increases. The shear stress is required to exhibit such trend by the imposed forcing term as discussed previously \citep{hamzehloo2021direct}. 
\par

\begin{figure*}
\centering
\includegraphics[width=4.95in]{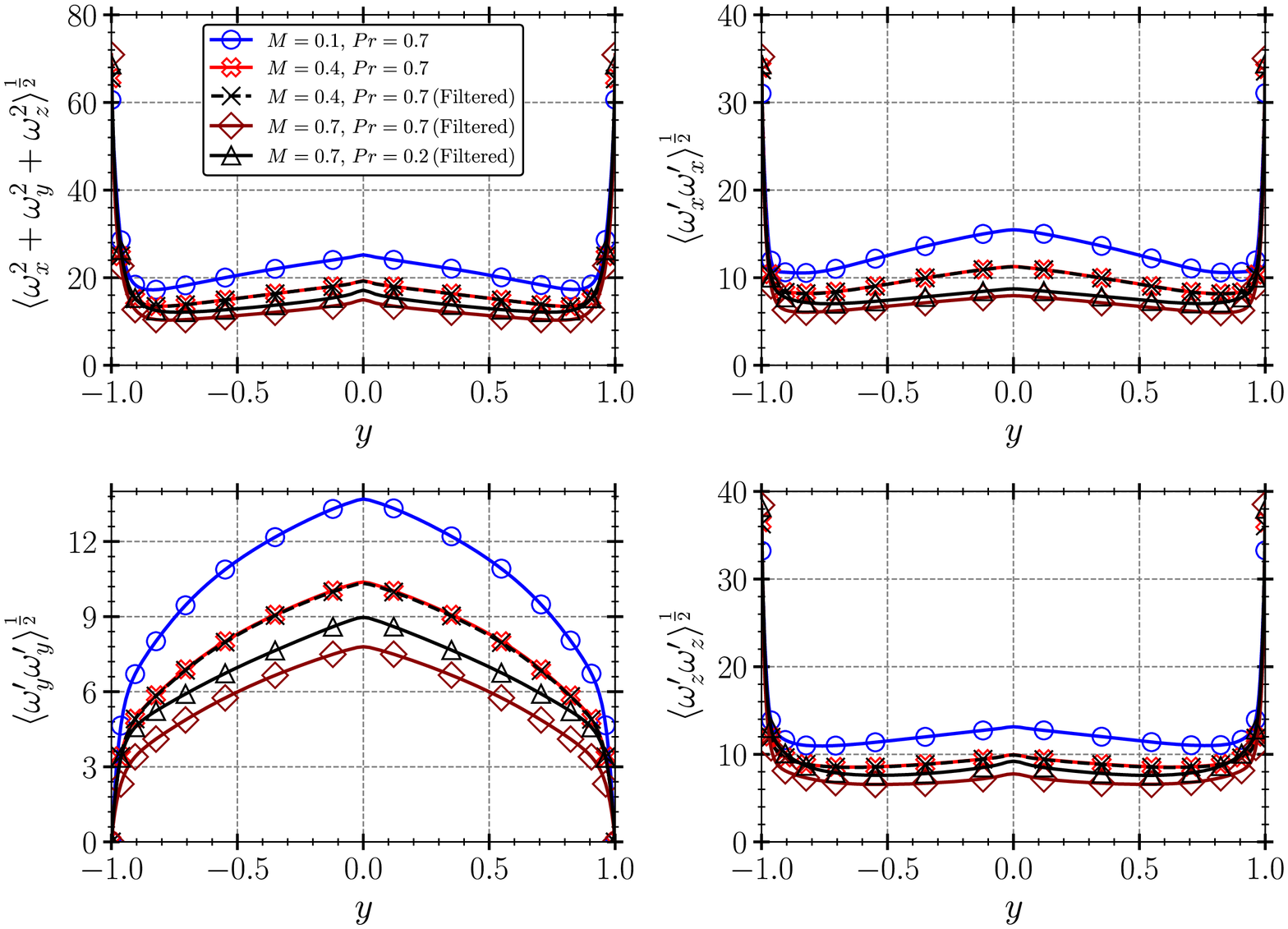}
\caption{\label{fig_vort} Profiles of the vorticity fluctuation and the components of the vorticity turbulent fluctuations.}
\end{figure*}

Figure \ref{fig_vort} shows the vorticity fluctuation ($\omega^{\prime}=\sqrt{\langle\omega_x^2+\omega_y^2+\omega_z^2\rangle}$) and the components of the vorticity turbulent fluctuations for different Mach and Prandtl numbers. With $Pr=0.7$, by increasing the Mach number, values of the vorticity fluctuations (total and all components) reduce noticeably. This trend is comparable to the relationship between the spanwise and wall-normal Reynolds stresses and the Mach number as seen in figure \ref{fig_mean}. The vorticity reduction is a sign of the three-dimensional structures weakening and turbulence stabilisation. Reducing the Prandtl number from $Pr=0.7$ to $0.2$ slightly increases the vorticity fluctuations. This is attributed to the significant increase in the streamwise Reynolds stress fluctuation as seen in figure \ref{fig_stress} and also an increase in the mean velocity  $\{ u \}$ as shown in figure \ref{fig_mean} and table \ref{table_cases}. The peak mean streamwise velocity increases by around $\sim 13.0 \%$  as the Prandtl number reduces.   
\par
As shown in figure \ref{fig_mean}, the counter-flow with $M=0.4$ forms a mildly compressible flow with $M{_t}_p\approx0.6$, hence the formation of discontinuities in the form of shocklets is limited. Figures \ref{fig_mean} and \ref{fig_stress} show that the cases with $M=0.4$ without and with the WENO filtering exhibit almost identical trends and table \ref{table_cases} shows less than $\sim 1 \%$ changes in their key flow quantities. This suggests that the filter-based shock capturing method is performing well and the grid resolution is fine enough.  

\subsection*{Shocklet Structure}

With the values of the mean and turbulent Mach numbers seen in figure \ref{fig_mean}, regions with instantaneous Mach numbers beyond unity are expected to form in the cases with $M=0.7$. For such transonic values, the formation of shocklets is possible. Shocklets can be associated with regions where the dilatation, defined as $\displaystyle \theta= \frac{\partial u} {\partial x} + \frac{\partial v} {\partial y} + \frac{\partial w} {\partial z}$, is lower than a negative threshold i.e. $\theta < - \zeta$ \citep{samtaney2001direct, wang2017shocklet}. A value of $\zeta=3\theta^{\prime}$, where $\theta^{\prime}$ denotes the dilatation fluctuation (root mean square of the dilatation magnitude) defined as $ \displaystyle \theta^{\prime} = \sqrt{\langle(\frac{\partial u} {\partial x})^2+(\frac{\partial v} {\partial y})^2+(\frac{\partial w} {\partial z})^2\rangle}$, has been used in compressible decaying turbulence problems to detect shocklets as regions with strong compression rates (in turbulent Mach number values in the range of $0.5\leq M_t \leq 1.0$)  \citep{samtaney2001direct, wang2017shocklet}. Also, $\zeta=\theta^{\prime}$ was used by \citep{samtaney2001direct} to visualise shocklets. Table \ref{table_dil_fluc} provides the bulk-averaged values of the dilatation fluctuation ($\theta^{\prime}_b$) for the counter-flows studied here. 
\par
\begin{figure*}[!h]
\centering
\includegraphics[width=4.5 in]{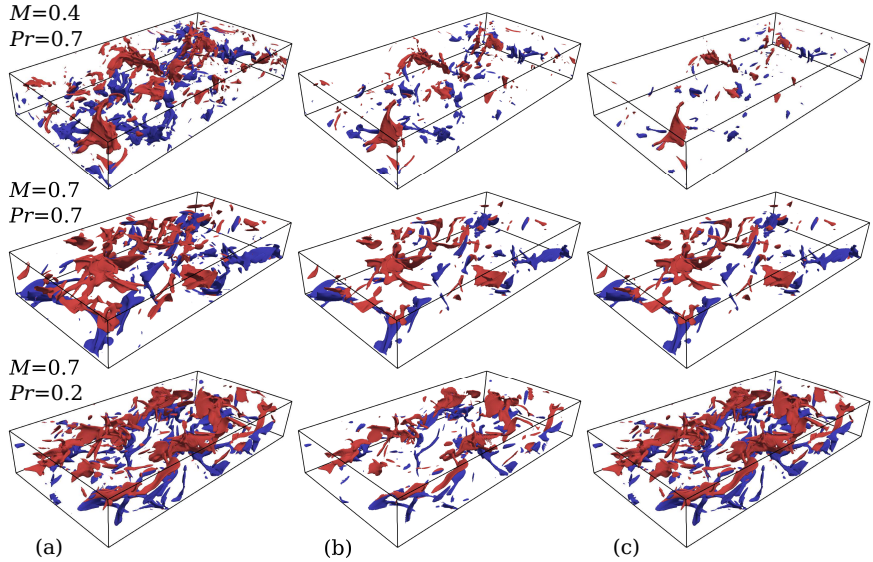}
\caption{\label{fig_shocklets} Iso-surfaces of the dilatation with iso-values of (a): $\theta_{iso}=2\theta^{\prime}_b$, (b): $\theta_{iso}=3\theta^{\prime}_b$ , and (c): $\theta_{iso}=3\theta^{\prime}_b|_{M=0.7,\, Pr=0.7}=5.655$. Red and blue colours show the flow directions in the positive (left to right) and negative streamwise directions, respectively.}
\end{figure*}

Following the literature, in the present study, in order to detect and visualise the shocklets, iso-surfaces of the dilatation with various threshold values (iso-values) of $\zeta=2\theta^{\prime}_b$, $3\theta^{\prime}_b$ and $3\theta^{\prime}_b|_{M=0.7,\, Pr=0.7}=5.655$ are used as shown in figure \ref{fig_shocklets} for counter-flows with $M\ge0.4$. The latter iso-value is used based on the bulk-averaged dilatation fluctuation of the case with $M=0.7$ and $Pr=0.2$ as a fixed reference with the aim of making a direct and relatively fairer comparison between the counter-flows studied. Based on what is seen from figure \ref{fig_shocklets}, and also observations reported in \citep{wang2017shocklet} a threshold value of $\zeta=3\theta^{\prime}_b$ can potentially be used to highlight the differences in high-compression regions of the flow (i.e. shocklet) as the compressibility increases in the counter-flow configuration. However, from subfigure (c) of figure \ref{fig_shocklets}, a fixed threshold may help better understand the flow differences. Specifically, as the compressibility increases, the number and size of the dilatation iso-surfaces increase significantly. It can be concluded that with $M=0.7$ the counter-flow can produce a shocklet-containing flow with significant numbers of heterogeneously-distributed highly irregular shocklets.                    

\begin{table}[!ht]
\centering
\caption{Bulk-averaged dilatation fluctuation.}
\begin{tabular}{c c c c }
\hline
 Case  &$M$& $Pr$& $\theta^{\prime}_b$\\
\hline
  1  & 0.1& 0.7& 0.128  \\
  3  & 0.4& 0.7& 1.374   \\
  4  & 0.7& 0.7& 1.885   \\
  5  & 0.7& 0.2& 2.808   \\
\hline
\end{tabular}
\label{table_dil_fluc}
\end{table}

\subsection*{Shocklet Quantification}

The total viscous dissipation can be divided into the solenoidal and dilatational components as $\displaystyle \epsilon ^{\mathcal{T}} = \epsilon ^{\mathcal{S}} + \epsilon ^{\mathcal{D}}$ \citep{sarkar1991analysis} where, the solenoidal ($\epsilon ^{\mathcal{S}}$) and dilatational (compressible) ($\epsilon ^{\mathcal{D}}$) dissipations are defined as $\displaystyle \epsilon ^{\mathcal{S}} = \frac{1}{Re}\Bigg[\mu \Bigg \langle \Big( \frac{dw}{dz} -  \frac{dv}{dz}\Big)^2 + \Big(\frac{du}{dz} -  \frac{dw}{dx} \Big)^2 + \Big( \frac{dv}{dx} -  \frac{du}{dy}\Big)^2 \Bigg \rangle \Bigg]$ and $\displaystyle \epsilon ^{\mathcal{D}} = \frac{4}{3 Re}\Bigg[\mu \Bigg \langle \frac{du}{dx} + \frac{dv}{dy} + \frac{dw}{dz}  \Bigg \rangle^2 \Bigg]$.

\begin{figure}[!h]
\centering
\includegraphics[width=2.5in]{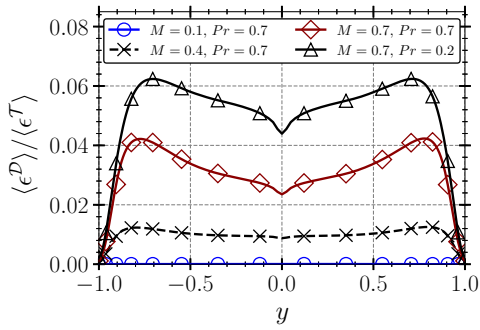}
\caption{\label{fig_dilat_disp} Profiles of the ratio of the dilatational dissipation over the total dissipation.}
\end{figure}

\noindent Energy dissipation through Mach number-induced changes on turbulent flow structures (i.e. shocklets) can be linked to the dilatational part of the dissipation \citep{sarkar1991analysis}. Figure \ref{fig_dilat_disp} shows the ratio of the dilatational dissipation to the total dissipation for the counter-flows studied here. The contribution of the dilatational part of the total viscous dissipation becomes significantly more important as the compressibility increases by having a higher Mach number and/or a lower Prandtl number. The peak of the ratio of the dilatational dissipation to the total dissipation occurs at around $|y|\approx0.75$ for all cases where the mean streamwise velocity and the mean Mach number also exhibit peak values as shown in figure \ref{fig_mean}. It is clear that $\epsilon ^{\mathcal{S}}$ is directly related to the vorticity which reduces as the Mach number increases. In fact, the solenoidal dissipation reduces slightly (not shown here) as the compressibility of the counter-flow increases. A higher compressibilty results in the formation of more shocklets (as shown in figure \ref{fig_shocklets}), and hence a relatively higher dilatational dissipation.

\par
\begin{figure}[!h]
\centering
\includegraphics[width=2.5in]{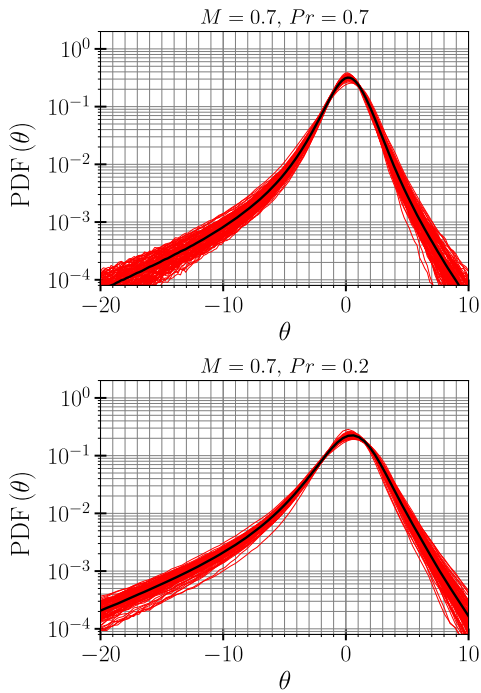}
\caption{\label{fig_pdfs_comp} Probability distributions of the dilatation over the entire domain for every $t=10$ intervals up to $t=1000$.}
\end{figure}

To provide a more quantitative analysis of the distribution of the shocklets, the Probability Density Function (PDF) of the dilation over the entire domain for time intervals of $t=10$ are plotted for the cases with $M=0.7$ in figure \ref{fig_pdfs_comp}. The red curves show the instantaneous PDFs and the thick black curves show their average in time. All PDF profiles are skewed towards the negative dilatation values, a trend expected due to the existence of nonlinear compression waves and shocklets in such compressible flows. Instantaneous PDF trends of figure \ref{fig_pdfs_comp} confirms that the dilatation, hence the number and strength of the shocklets, exhibits a significant fluctuating transient behaviour. However, as expected, the counter-flow with $Pr=0.2$ exhibits a more negatively skewed profile of the time-averaged dilatation PDF consistent with its higher compressibility level.

\section*{CONCLUSIONS}

\par
Direct numerical simulations of shocklet-containing turbulent flows were conducted using a new counter-flow channel configuration introduced previously by the current authors \citep{hamzehloo2021direct}. The simulations were performed for Mach number values of $M=0.1, 0.4$ and $0.7$ with a Reynolds number of $Re=400$. A case with a Prandtl number of $Pr=0.2$ (reduced from the default value of $Pr=0.7$) and $M=0.7$ was also studied in an attempt to boost the compressibility.      

\par
It was found that a Mach number as low as $M=0.7$ could produce fluctuating and mean Mach numbers close to $0.7$ over a considerable length of the width of the counter-flow channel. Such values produced instantaneous supersonic velocities that formed relatively strong transient shocklets. Additionally, fluctuating and mean Mach numbers above $0.95$ were achieved by reducing the Prandtl number from $Pr=0.7$ to $0.2$. Despite the presence of shocklets the contribution of dilatational dissipation to the total dissipation was found to be small (6\% or less).

\par
Overall, the counter-flow configuration was found to be able to produce highly turbulent flows with embedded shocklets for a relatively modest Mach number. Therefore, the configuration provides a useful framework to study some of the fundamental physics associated with shock-turbulence interactions and could be considerably beneficial to the development of compressible subgrid scale turbulence models as well as improved representations of compressibility in Reynolds-averaged Navier-Stokes models.  


\section*{ACKNOWLEDGMENTS}

Dr Arash Hamzehloo was funded by the UK Turbulence Consortium (EPSRC grant EP/R029326/1). Dr David J Lusher was funded by EPSRC grant EP/L015382/1. The authors acknowledge the use of the Cambridge Tier-2 system operated by the University of Cambridge Research Computing Service under an EPSRC Tier-2 capital grant (EP/P020259/1). The flow solver OpenSBLI is available at \href{https://opensbli.github.io}{https://opensbli.github.io}.

\bibliographystyle{tsfp}
\bibliography{tsfp}

\begin{thebibliography}{11}
\expandafter\ifx\csname natexlab\endcsname\relax\def\natexlab#1{#1}\fi

\bibitem[Forliti {\em et~al.\/}(2005)Forliti, Tang \&
  Strykowski]{forliti2005experimental}
Forliti, D.J., Tang, B.A. \& Strykowski, P.J. 2005 An experimental
  investigation of planar countercurrent turbulent shear layers. {\em Journal
  of Fluid Mechanics\/} {\bf 530}, 241.

\bibitem[Freund {\em et~al.\/}(2000)Freund, Lele \&
  Moin]{freund2000compressibility}
Freund, J.B., Lele, S.K. \& Moin, P. 2000 Compressibility effects in a
  turbulent annular mixing layer. {P}art 1. {T}urbulence and growth rate. {\em
  Journal of Fluid Mechanics\/} {\bf 421}, 229--267.

\bibitem[Hamzehloo {\em et~al.\/}(2021)Hamzehloo, Lusher, Laizet \&
  Sandham]{hamzehloo2021direct}
Hamzehloo, A., Lusher, D.J., Laizet, S. \& Sandham, N.D. 2021 Direct numerical
  simulation of compressible turbulence in a counter-flow channel
  configuration. {\em Physical Review Fluids\/} {\bf 6}~(9), 094603.

\bibitem[Humphrey \& Li(1981)]{humphrey1981tilting}
Humphrey, J.A.C. \& Li, S. 1981 Tilting, stretching, pairing and collapse of
  vortex structures in confined counter-current flow. {\em Journal of Fluids
  Engineering\/} {\bf 103}~(3).

\bibitem[Lusher {\em et~al.\/}(2021)Lusher, Jammy \&
  Sandham]{lusher2021opensbli}
Lusher, D.J., Jammy, S.P. \& Sandham, N.D. 2021 {O}pen{SBLI}: Automated
  code-generation for heterogeneous computing architectures applied to
  compressible fluid dynamics on structured grids. {\em Computer Physics
  Communications\/} p. 108063.

\bibitem[Samtaney {\em et~al.\/}(2001)Samtaney, Pullin \&
  Kosovi{\'c}]{samtaney2001direct}
Samtaney, R., Pullin, D.I. \& Kosovi{\'c}, B. 2001 Direct numerical simulation
  of decaying compressible turbulence and shocklet statistics. {\em Physics of
  Fluids\/} {\bf 13}~(5), 1415--1430.

\bibitem[Sarkar {\em et~al.\/}(1991)Sarkar, Erlebacher, Hussaini \&
  Kreiss]{sarkar1991analysis}
Sarkar, S., Erlebacher, G., Hussaini, M.Y. \& Kreiss, H.O. 1991 The analysis
  and modelling of dilatational terms in compressible turbulence. {\em Journal
  of Fluid Mechanics\/} {\bf 227}, 473--493.

\bibitem[Strykowski \& Wilcoxon(1993)]{strykowski1993mixing}
Strykowski, P.J. \& Wilcoxon, R.K. 1993 Mixing enhancement due to global
  oscillations in jets with annular counterflow. {\em AIAA Journal\/} {\bf
  31}~(3), 564--570.

\bibitem[Vreman {\em et~al.\/}(1996)Vreman, Sandham \&
  Luo]{vreman1996compressible}
Vreman, A.W., Sandham, N.D. \& Luo, K.H. 1996 Compressible mixing layer growth
  rate and turbulence characteristics. {\em Journal of Fluid Mechanics\/} {\bf
  320}, 235--258.

\bibitem[Wang {\em et~al.\/}(2017)Wang, Gotoh \& Watanabe]{wang2017shocklet}
Wang, J., Gotoh, T. \& Watanabe, T. 2017 Shocklet statistics in compressible
  isotropic turbulence. {\em Physical Review Fluids\/} {\bf 2}~(2), 023401.

\bibitem[Yee \& Sjögreen(2018)]{YEE2018331}
Yee, H.C. \& Sjögreen, B. 2018 Recent developments in accuracy and stability
  improvement of nonlinear filter methods for dns and les of compressible
  flows. {\em Computers \& Fluids\/} {\bf 169}, 331--348.

\end{thebibliography}

\end{document}